\documentclass[preprint,aps,pra,showpacs,floats,superscriptaddress]{revtex4}
\usepackage[dvips]{graphicx}
\usepackage{amsmath,bm}
\usepackage{psfrag}

\begin{document}

\title{Comparative study of a solid film dewetting in an attractive substrate potentials
with the exponential and the algebraic decay}

\author{M.~Khenner}
\affiliation{Department of Mathematics, State University of New
York at Buffalo, Buffalo, NY 14260, USA}

\newcommand{\Section}[1]{\setcounter{equation}{0} \section{#1}}
\newcommand{\rf}[1]{(\ref{#1})}
\newcommand{\beq}[1]{ \begin{equation}\label{#1} }
\newcommand{\eeq}{\end{equation} }

\begin{abstract}
We compare dewetting characteristics of a thin nonwetting solid film in the absence of stress, for
two models of a wetting potential: the exponential and the algebraic. 
The exponential model is a one-parameter ($r$) model, and the algebraic model is a two-parameter ($r,\; m$) 
model, where $r$ is the ratio of the characteristic wetting length to the height of the unperturbed film,
and $m$ is the exponent of $h$ (film height) in a smooth function that interpolates the system's surface 
energy above and below the film-substrate interface at $z=0$. The exponential model
gives monotonically decreasing (with $h$) wetting chemical potential, while
this dependence is monotonic only for the $m=1$ case of the algebraic model.
Linear stability analysis of the planar equilibrium surface is performed. Simulations of the surface dynamics in the strongly nonlinear regime
(large deviations from the planar equilibrium) and for large surface energy anisotropies demonstrate that 
for any $m$ the film is less prone to dewetting
when it is governed by the algebraic model. Quasiequilibrium states similar to the one found in the exponential
model \cite{Dwt2} exist in the algebraic model as well, and the film morphologies are similar.

\end{abstract}

\date{\today}
\maketitle

\begin{center}
{\bf I. INTRODUCTION}
\end{center}

Dewetting of lattice-matched ultrathin solid films (such as the sub-10 nm Si film on the SiO$_2$ substrate) was recently observed in 
experiments at temperatures around 800$^\circ$C \cite{YZSLLHL,SECS}. 
Presumably, the cause for film dewetting is a long-range, attractive film-substrate interaction 
(also called wetting interaction) which amplifies perturbations of the planar film surface and makes the film height 
decrease locally until the surface reaches the substrate, resulting in the formation of an array of islands. 
At this most general level of description
dewetting of solid films is similar to dewetting of liquid films (which has been studied for many years \cite{Oron,SHJ}), the only difference is the nature of the mass transport, i.e. the thermally activated surface diffusion of adatoms in the former case vs. the fluid flow in the latter case. There is, however, two determinative reasons of as to why the dynamics of dewetting in these systems is qualitatively different. One reason is the nonzero (and generally, strong) anisotropy of the solid film surface energy (tension) which is not present in liquids. 
As has been shown by the author in Refs. \cite{Dwt1,Dwt2}, faceting of the surface due to strong anisotropy  
opposes the tendency of the film to dewet.
Another reason is ``geometrical", meaning that a planar
surface of the as-deposited solid film may feature local defects of arbitrary shape protruding arbitrarily deep into 
the film (i.e., the pinholes). Since the attractive substrate potential decreases with the film height, its influence is stronger  on deep pinholes, which therefore dewet faster. In contrast to shallow pinholes the morphology
of the tip is often different from the morphology of other parts of the surface, i.e. the surface away from the tip may 
undergo formation of a hill-and-valley structure due to faceting \cite{Dwt1,Dwt2}.

These and other differences as well as importance to technologies such as the design and manufacture of solid thin-film devices, make dewetting of solid films a process worth studying. 
In Refs. \cite{Dwt1,Dwt2} analytical and computational studies are performed of the two-dimensional PDE-based model,
which incorporates the two-layer wetting potential with the exponential decay.
Previously, Golovin \textit{et al.} and other authors \cite{GDV}-\cite{Chiu} 
studied similar models in the context of quantum dots self-assembly. 
Note that the two-layer potential model is appropriate for ultrathin solid films, while for thicker films
the van der Waals potential has been shown to be important \cite{SZ}.
In this paper the model of Refs. \cite{Dwt1,Dwt2} 
is extended to the case of the two-layer wetting potential with 
a variable-rate algebraic decay, and comparisons of the two situations are performed. The models are studied using the 
linear stability analysis, as well as the computations of the arbitrary deviation/slope surface dynamics.

\begin{center}
{\bf II. PROBLEM FORMULATION}
\end{center}

The governing equation for the free one-dimensional (1D) surface $z=h(x,t)$, evolving by surface diffusion, has the form
\begin{equation}
h_t = \frac{\Omega D \nu}{kT}\frac{\partial}{\partial x}\left((1+h_x^2)^{-1/2}\frac{\partial \mu}{\partial x}\right),
\label{1.1}
\end{equation}
where $h$ is the height of the film above the substrate, $\Omega$ is the atomic volume, 
$D$ the adatoms diffusivity, $\nu$ the adatoms surface density, $k$ the Boltzmann constant,
$T$ the absolute temperature, and $\mu = \mu^{(\kappa)}+\mu^{(w)}$ the surface chemical potential.
Here $\mu^{(\kappa)}$ is the regular contribution due to the surface mean curvature $\kappa$ \cite{MULLINS5759}.
Also $(1+h_x^2)^{-1/2}=\cos{\theta}$,
where $\theta$ is the angle that the unit surface normal makes with the [01] crystalline direction, along which is
the $z$-axis.
(The $x$-axis is along the [10] direction.)
Thus $\theta$ measures the orientation of the surface with respect to the underlying crystal structure.
Note throughout the paper the subscripts $x, t, s, u$ and $\theta$ denote differentiation.

The wetting chemical potential 
\begin{equation}
\mu^{(w)} = \Omega \left(1+h_x^2\right)^{-1/2}\frac{\partial \gamma}{\partial h},
\label{1.4e}
\end{equation}
where $\gamma$ is the height-dependent surface energy of the film-substrate interface. 
In the two-layer exponential wetting model \cite{CG}
\begin{equation}
\gamma(h,\theta) = \gamma^{(f)}(\theta) + \left(\gamma_S-\gamma^{(f)}(\theta)\right)\exp{\left(-h/\ell\right)},\quad h>0.
\label{1.4f}
\end{equation}
In the two-layer algebraic wetting model \cite{BrianWet}
\begin{equation}
\gamma(h,\theta) = \frac{1}{2}\left(\gamma^{(f)}(\theta) + \gamma_S\right)+\frac{1}{2}\left(\gamma^{(f)}(\theta)-\gamma_S\right)
f(h/\ell),\quad \lim_{h\rightarrow \infty} f(h/\ell)=1,\quad \lim_{h\rightarrow -\infty} f(h/\ell)=-1.
\label{1.4f1}
\end{equation}
Here $\gamma_S=const.$ is the surface energy density of the substrate in the absence
of the film, and $\ell$ is the characteristic wetting length. 
$\gamma^{(f)}(\theta)$ is the energy of the film surface, assumed
strongly anisotropic. 
In the exponential model $\gamma(h,\theta) \rightarrow \gamma^{(f)}(\theta)$ 
as $h\rightarrow \infty$, and $\gamma(h,\theta) \rightarrow \gamma_S$ as $h\rightarrow 0$.
In the algebraic model $f(h/\ell)$ is such that (i) the correct surface energies, $\gamma^{(f)}(\theta)$ and $\gamma_S$, 
are recovered as $h\rightarrow \pm \infty$, and (ii) approach to the limiting value +1 as $h\rightarrow \infty$ is an algebraic power. (Of course, negative film height has no physical meaning, thus formally in the substrate domain $h$ 
must be replaced by $z$ in Eq. \rf{1.4f1}.)
The suitable generic form is \cite{BrianWet,KF}:
\begin{equation}
f(h/\ell)=\frac{2}{\pi}\mbox{arctan}\left[\left(\frac{h}{\ell}\right)^m\right],\quad m=1,3,5,\ldots
\label{1.4f2}
\end{equation}
which has the expansion
\begin{equation}
f(h/\ell)=1-\frac{2}{\pi}(h/\ell)^{-m}+\ldots \quad \mbox{as}\;\; h\rightarrow \infty.
\label{1.4f3}
\end{equation}
Note that in the limit $h\rightarrow 0$ the exponential and the algebraic models give 
$\gamma=\gamma_S$ and $\gamma=\left(\gamma^{(f)}(\theta)+\gamma_S\right)/2$, respectively.
These results follow from the `one-sided' 
(`two-sided') nature of the the corresponding boundary layer models for the smooth transition in surface energy
above (across) the substrate surface $z=0$, over a small length scale $\ell$.

$\gamma^{(f)}(\theta)$ is taken in the form 
\begin{equation}
\gamma^{(f)}(\theta) = \gamma_0 (1+\epsilon_\gamma \cos{4\theta}) + \frac{\delta}{2}\kappa^2 \equiv 
\gamma_p(\theta) + \frac{\delta}{2}\kappa^2,
\label{1.2}
\end{equation}
where $\gamma_0$ is the mean value of the film surface energy
in the absence of the substrate
potential (equivalently, the surface energy of a very thick film), $\epsilon_\gamma$ determines the degree of anisotropy,
and $\delta$ is the small non-negative regularization parameter having units of energy. 
The $\delta$-term in Eq. \rf{1.2} makes the evolution equation \rf{1.1} mathematically well-posed for strong 
anisotropy \cite{AG} - \cite{SG}. 
(The anisotropy is weak when $0<\epsilon_\gamma<1/15$ and strong when $\epsilon_\gamma\ge 1/15$.
$\delta=0$ in the former case. In the latter case the polar plot of $\gamma^{(f)}(\theta)$ has cusps at the orientations
that are missing from the equilibrium Wulff shape and the surface stiffness 
$\gamma^{(f)}+\gamma^{(f)}_{\theta\theta}$ is negative
at these orientations \cite{HERRING,HERRING1}. Thus the evolution equation is ill-posed unless regularized 
\cite{AG,CGP}.)
The form \rf{1.2} assumes that the surface energy is maximum in the [01] direction.
With the regularization in place, the curvature contribution to the chemical potential has the standard form 
\begin{equation}
\mu^{(\kappa)} = 
\Omega \left[(\gamma+\gamma_{\theta \theta})\kappa-\delta\left(\frac{\kappa^3}{2}+\kappa_{ss}\right)\right], 
\label{1.4b}
\end{equation}
where $\kappa = -h_{xx}(1+h_x^2)^{-3/2}$, $s$ is the arclength along the surface 
[$\partial/\partial s = (\cos{\theta})\partial/\partial x$]
and the expressions for $\gamma(h,\theta)$ read 
\begin{equation}
\mbox{Exponential model}:\quad \gamma(h,\theta) = \gamma_p(\theta) + \left(\gamma_S-\gamma_p(\theta)\right)\exp{\left(-h/\ell\right)},
\label{1.4fp}
\end{equation}
\begin{equation}
\mbox{Algebraic model}:\quad \gamma(h,\theta) = \frac{1}{2}\left(\gamma_p(\theta) + \gamma_S\right)+\frac{1}{2}\left(\gamma_p(\theta)-\gamma_S\right)
f(h/\ell),
\label{1.4f1p}
\end{equation}
with $\gamma_p(\theta)$ stated in Eq. \rf{1.2}. 
By using Eqs. \rf{1.4fp} and \rf{1.4f1p} instead of Eqs. \rf{1.4f} and \rf{1.4f1} we disregard the contribution of
the wetting terms (exponential or inverse tangent) to the regularization in Eq. \rf{1.4b}.
Similarly, by using Eqs. \rf{1.4fp} and \rf{1.4f1p} in Eq. \rf{1.4e},
we disregard the contribution of the regularization term $\delta \kappa^2/2$ to $\mu^{(w)}$. 
(See Refs. \cite{Dwt1,Dwt2} for the justification of this approach.)

Using the height of the planar unperturbed film, $h_0$, as the length scale, the nondimensional expressions for the
chemical potentials read:\\
Exponential model:
\begin{subequations}
\begin{eqnarray}
\mu^{(\kappa)} &=& 
\left(\bar \gamma_p(\theta)+\frac{\partial^2\bar \gamma_p}{\partial \theta^2}\right)
\left(1-\exp{\left(-h/r\right)}\right)\kappa + \Gamma\exp{\left(-h/r\right)}\kappa - \Delta\left(\frac{\kappa^3}{2}+\kappa_{ss}\right),\\
\mu^{(w)} &=& \left(\bar \gamma_p(\theta) - \Gamma\right)\frac{\exp{\left(-h/r\right)}}{r} \cos{\theta}.
\label{ndim_eq}
\end{eqnarray}
\end{subequations}
Algebraic model:
\begin{subequations}
\begin{eqnarray}
\mu^{(\kappa)} &=& 
\frac{1}{2}\left(\bar \gamma_p(\theta)+\frac{\partial^2\bar \gamma_p}{\partial \theta^2}\right)
\left(1+f\left(h/r\right)\right)\kappa + \frac{\Gamma}{2}\left(1-f\left(h/r\right)\right)\kappa - \Delta\left(\frac{\kappa^3}{2}+\kappa_{ss}\right),\\
\mu^{(w)} &=& \frac{1}{2}\left(\bar \gamma_p(\theta) - \Gamma\right)\frac{df}{dh} \cos{\theta},
\label{ndim_eq1}
\end{eqnarray}
\end{subequations}
where 
\begin{equation}
f(h/r)=\frac{2}{\pi}\mbox{arctan}\left[\left(\frac{h}{r}\right)^m\right],\quad m=1,3,5,\ldots ,\quad 
\bar \gamma_p(\theta) = 1+\epsilon_\gamma \cos{4\theta}.
\label{1.4f2nd}
\end{equation}
Also, $r = \ell/h_0$ is the ratio of the characteristic wetting length to the unperturbed film height,
$\Gamma = \gamma_S/\gamma_0$ is the ratio of the mean surface energy of the film  
to the substrate surface energy, and $\Delta = \delta/(\gamma_0h_0^2)$
is the non-dimensional regularization parameter.
Figures 1 and 2 show $\mu^{(w)}$ for both models. Note that for the algebraic model $\mu^{(w)}\sim 1/h^{m+1}$ for $h\gg 1$,
as follows from Eqs. \rf{1.4e}, \rf{1.4f1} and \rf{1.4f3}. 
\begin{figure}[!h]
\includegraphics[width=4.5in]{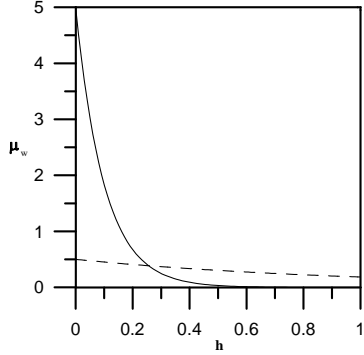}
\vspace{-8.0cm}
\caption{The reduced nondimensional wetting chemical potential $\mu^{(w)}=\exp{(-h/r)}/2r$.
This formula is obtained when $\bar \gamma_p$ is taken isotropic, $h_x$ is taken zero
and $\Gamma = 0.5$
in Eq. \rf{ndim_eq}. 
Solid line: $r=0.1$; dash line: $r=1$.} \label{fig:1}
\end{figure}
\begin{figure}[!t]
\includegraphics[width=3.5in]{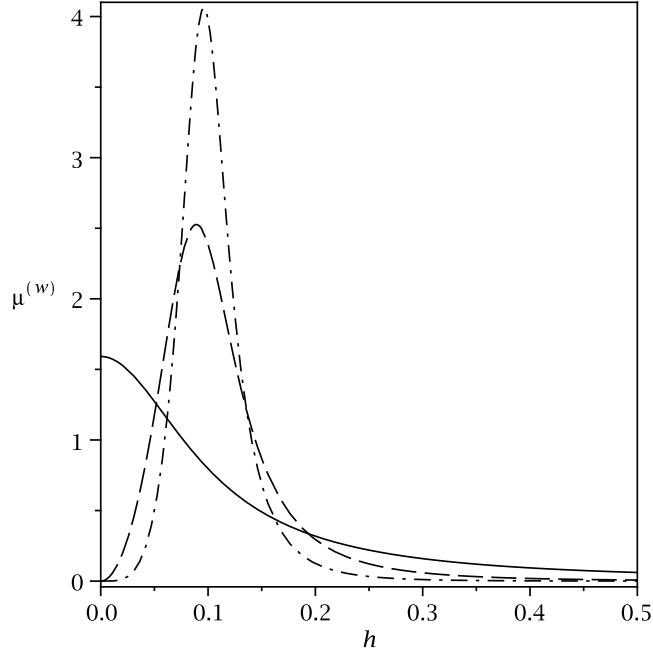}
\caption{The reduced nondimensional wetting chemical potential 
$\mu^{(w)}=(1/4)df/dh$, where $f$ is given by Eq. \rf{1.4f2nd}.
This formula is obtained when $\bar \gamma_p$ is taken isotropic, $h_x$ is taken zero
and $\Gamma = 0.5$ in Eq. \rf{ndim_eq1}. Also $r=0.1$.
Solid line: $m=1$; dash line: $m=3$; dash-dot line: $m=5$.} \label{fig:2}
\end{figure}

Now using
$h_0^2/D$ as the time scale, and the small-slope expansion in powers of $\epsilon = |\partial/\partial x| \ll 1$, the asymptotic nondimensional evolution equation \rf{1.1} reads:
\begin{equation}
h_t = B\frac{\partial}{\partial x}\left(P^{(1)}_\kappa-\Delta P^{(2)}_\kappa+P_w\right),
\label{1.5}
\end{equation}
where $B = \Omega^2\nu \gamma_0/(kTh_0^2)$ is the Mullins coefficient and $P^{(2)}_\kappa = -h_{xxxxx}$.

In the exponential model, the terms $P^{(1)}_\kappa$ and $P_w$ read:
\begin{equation}
P^{(1)}_\kappa = \Lambda_1h_{xxx}+\Lambda_2h_{xx}^2h_x+\Lambda_3h_{xxx}h_x^2-\exp{\left(-h/r\right)}\left[\left(\Gamma+\Lambda_1\right)h_{xxx}+o.t.\right],
\label{1.8a}
\end{equation}
\begin{equation}
P_w = \frac{\exp{\left(-h/r\right)}}{r}\left[a_2h_{xx}h_x\left(1-5h_x^2\right)+
r^{-1}h_x\left(a_1+\left(2a_3-3a_1\right)h_x^2+a_4h_x^4\right)\right],
\label{1.8c}
\end{equation}
where $\Lambda_1 = 15\epsilon_\gamma - 1,\ \Lambda_2 = 3-285 \epsilon_\gamma,\
\Lambda_3 = 2-150 \epsilon_\gamma,\ 
a_1 = \Gamma - 1 - \epsilon_\gamma,\ a_2 = \Gamma - 1 - 17\epsilon_\gamma,\ a_3 = \Gamma - 1 + 3\epsilon_\gamma,\ a_4 = \Gamma - 1 - 25\epsilon_\gamma$.  
Notation \textit{o.t.} (meaning \textit{other terms})  in the second part of Eq. \rf{1.8a} (which is proportional to the exponent and which
stems from wetting interaction), and in the following Eqs. \rf{1.8aa}, \rf{1.8ca} stands for many omitted terms that do not contribute to linear stability. Note that the non-negative $\Lambda_1$ signals that the surface energy anisotropy is strong.

In the algebraic model the terms $P^{(1)}_\kappa$ and $P_w$ read:
\begin{eqnarray}
P^{(1)}_\kappa &=& \frac{h_{xxx}}{2\pi\left(1+(h/r)^{2m}\right)}\left[2\left\{\Gamma+\Lambda_1+a_5(h/r)^{2m}\right\}\mbox{arctan}\left[\left(h/r\right)^m\right]- \right. \nonumber \\ 
&& \left.\pi a_6\left(1+(h/r)^{2m}\right)\right]+o.t.,
\label{1.8aa}
\end{eqnarray}
\begin{equation}
P_w = \frac{m}{\pi}a_1\frac{\left(h/r\right)^m h_x}{h^2\left(1+\left(h/r\right)^{2m}\right)^2}\left[1-m+(1+m)
(h/r)^{2m}\right]+o.t.,
\label{1.8ca}
\end{equation}
where 
$a_5 = \Gamma - 1 +15\epsilon_\gamma$ and 
$a_6 = \Gamma +1 -15\epsilon_\gamma$.

For computations of the surface evolution (Section IV) we use the 
parametric equations and the marker particle method. The 1D surface is specified as
$\Upsilon(x(u,t),z(u,t))$, where
$u$ is the parameter. $x$ and $z$  represent the coordinates of 
a marker particle on a surface, which are governed by two coupled nondimensional PDEs \cite{Sethian1}-\cite{HLS}:
\begin{subequations}
\label{base_eq}
\begin{eqnarray}
x_t &=& V \frac{1}{g}z_u,\\
z_t &=& - V \frac{1}{g}x_u.
\end{eqnarray}
\end{subequations}
Here 
$V=B\left(\mu^{(\kappa)}_{ss} + \mu^{(w)}_{ss}\right)$ is the normal velocity of the surface,
and $g=ds/du=\sqrt{x_u^2+z_u^2}$ is the metric function. It can be easily shown that Eqs. \rf{base_eq} are 
equivalent to dimensionless 
Eq. \rf{1.1} when the surface is non-overhanging (a graph of $h=h(x)$ at all times). (Note that in this case
$u\equiv x$ and $\partial/\partial s = g^{-1}\partial/\partial u = (1+h_x^2)^{-1/2}\partial/\partial x$.)
When the surface develops steep slope, the accurate computation using Eq. \rf{1.1} requires a fine grid, 
and when the surface overhangs, Eq. \rf{1.1} does not make sense. Eqs. \rf{base_eq} and the marker particle method
allow to circumvent these problems, and thus this combination is
preferred for computation of evolving general surfaces.

\bigskip
\begin{center}
{\bf III. LINEAR STABILITY ANALYSIS OF THE PLANAR SURFACE}
\end{center}

\begin{center}
{\bf A. Exponential Model}
\end{center}

We assume strong anisotropy and linearize
Eq. \rf{1.5} about the equilibrium $h=1$. For the perturbation $\xi(x,t)$ we obtain
\begin{equation}
\xi_t = B\left(\Lambda_1\xi_{xxxx}+\Delta\xi_{xxxxxx}+\exp{\left(-1/r\right)}\left[r^{-2}a_1\xi_{xx}-(\Gamma+\Lambda_1)\xi_{xxxx}\right]\right).
\label{1.11}
\end{equation}
Taking  $\xi = e^{ikx+\omega t}$ gives
\begin{equation}
\omega(k) = B\left[\left(\Lambda_1-\exp{\left(-1/r\right)}\left(\Gamma+\Lambda_1\right)\right)k^4-\Delta k^6-
\exp{\left(-1/r\right)}r^{-2}a_1k^2\right].
\label{1.12}
\end{equation}
Note that taking the limit as $r\rightarrow 0$ in Eq. \rf{1.12} recovers the dispersion relation in the absence of
wetting interaction with the substrate, $\omega(k) = B\left[\Lambda_1 k^4-\Delta k^6\right]$.
It follows from Eq. \rf{1.12} that 
the equilibrium surface is unstable ($\omega(k)>0$) to perturbations with the wavenumbers $0< k < k_c$, where
\begin{eqnarray}
k_c^2 &=& (2\Delta)^{-1}\left[\Lambda_1-\exp{\left(-1/r\right)}\left(\Gamma+\Lambda_1\right)+ \right.\nonumber \\
&& \left. 
\left(\left(\Lambda_1-\exp{\left(-1/r\right)}\left(\Gamma+\Lambda_1\right)\right)^2-4\Delta\exp{\left(-1/r\right)}r^{-2}a_1\right)^{1/2}\right].
\label{1.12a}
\end{eqnarray}
Note that the radical at the right-hand side of Eq. \rf{1.12a} always exists when $a_1<0$,
which turns out to be the necessary condition for a nonwetting film \cite{Dwt1}.
Fig. 3 shows the sketch of $\omega(k)$.
It is interesting that the $k^2$-term in Eq. \rf{1.12} coming from $P_w$ (Eq. \rf{1.8c}) makes the film less stable for 
$a_1<0$, but the wetting potential contribution to $P^{(1)}_\kappa$ in Eq. \rf{1.8a} makes the surface more
stable since the corresponding $k^4$-term is negative in Eq. \rf{1.12}. Clearly, due to negative exponent this 
stabilizing influence is small when $r$ is small.
\begin{figure}[!t]
\includegraphics[width=3.5in]{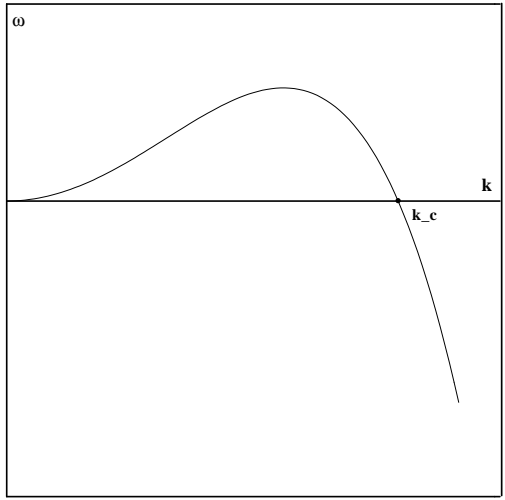}
\vspace{-2.0cm}
\caption{Sketch of the linear growth rate $\omega(k)$. Perturbations with wavenumbers
$0< k < k_c$ are unstable and may grow nonlinearly until the film ruptures.} \label{fig:3}
\end{figure}

\begin{center}
{\bf B.  Algebraic Model}
\end{center}

Eq. \rf{1.5} gives
\begin{eqnarray}
\omega(k) &=& B\left[\frac{2\left\{\Gamma+\Lambda_1+a_5 r^{-2m}\right\}A_0-\pi a_6\left(1+r^{-2m}\right)}{2\pi\left(1+r^{-2m}\right)}k^4 - \Delta k^6- \right.\nonumber \\
&& \left.\frac{m\left[1-m+(1+m)
r^{-2m}\right]}{\pi r^m \left(1+r^{-2m}\right)^2}a_1 k^2\right],
\label{1.13}
\end{eqnarray}
where $A_0=\mbox{arctan}\left[\left(1/r\right)^m\right]$. 
%Attempting to recover the dispersion relation in the absence of
%wetting interaction with the substrate, one has to 
%take the limit as $r\rightarrow \infty$ in Eq. \rf{1.13} and set $\Gamma=0$ in $a_6$. This results in 
%$\omega(k) = B\left[\Lambda_1 k^4/2-\Delta k^6\right]$, i.e. there is a factor of two difference in the first term. 

The cut-off wavenumber is compared in Fig. 4 
for both models and the three values of $m$.
For $m=1$, $k_c$ tends to zero asymptotically, while for $m=3,\ 5$ it becomes zero at $r$ slightly larger than one.
Thus the $m=1$  case is qualitatively similar to the exponential model. 
Comparing the $m=1$ case to the exponential model, 
it can be seen that for $r < 0.5$ the surface is more stable in the former case, and less stable for $r >0.5$.
Comparing the $m=3,\ 5$ cases to the exponential model,  it can be seen that for small values or $r$ the interval of instability is the same for both models,
for intermediate values of $r$ the interval is larger for the algebraic model, and 
for $r>\sim 1$ the surface governed by the algebraic model is absolutely stable, while there is still a narrow
interval of long-wave instability in the exponential model. 
\begin{figure}[!t]
\includegraphics[width=3.5in]{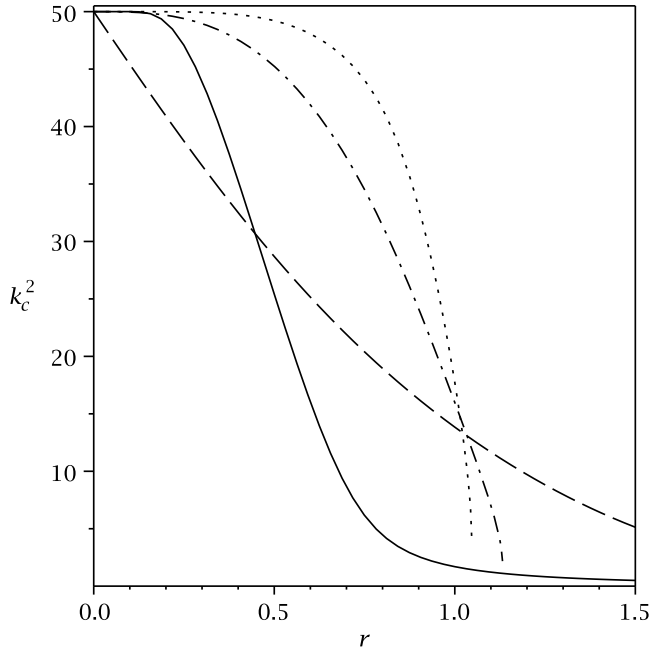}
\caption{Plots of the square of the critical wavenumber vs. $r$.
$\Gamma = 0.5,\; \epsilon_\gamma = 1/12,\; \Delta = 0.005$. 
Solid line: exponential model;  dash line: algebraic model with $m=1$; dash-dot line: algebraic model with $m=3$; 
dot line: algebraic model with $m=5$. (Abrupt termination of the dash-dot and dot lines is the artifact of the 
plotting software. We confirmed that these lines continue to intersection with the $r$-axis.)
} \label{fig:4}
\end{figure}
%
%
%\begin{figure}[!t]
%\includegraphics[width=3.5in]{fig5_kc2_comp1.eps}
%\caption{Comparison of the $m=1$ case of the algebraic model (dashed line) to the
%exponential model (solid line).} \label{fig:6}
%\end{figure}
%
%
%\begin{figure}[!t]
%\includegraphics[width=3.5in]{fig5_kc2_comp2.eps}
%\caption{Comparison of the $m=3,\ 5$ cases of the algebraic model (dashed and dash-dot lines, respectively) to the
%exponential model (solid line).} \label{fig:7}
%\end{figure}
%

\bigskip
\begin{center}
{\bf IV. NUMERICAL RESULTS FOR THE LARGE-AMPLITUDE INITIAL DEFORMATION (PINHOLE DEFECT)}
\end{center}

In this section the parameters are chosen as follows: $r=0.1,\; \Gamma = 0.5,\; \epsilon_\gamma = 1/12,\; \Delta = 0.005$.
Following the method of lines approach, Eqs. \rf{base_eq} are discretized in the parameter $u$ using second-order finite differences and the time-stepping is performed by the implicit Runge-Kutta solver RADAU \cite{RADAU}.
Initially $u\equiv x$, but periodically (usually after every few tens of the time steps) the surface is reparametrized
so that $u$ becomes the arclength, and the positions of the marker particles are recomputed accordingly.
This prevents marker particles from coming too close or too far apart in the course of the surface evolution.

We compute the dynamic morphology and its rate of evolution towards either film rupture or the quasiequilibrium
state, which is characterized by the coarsening in time hill-and-valley structure at the both sides of the residual
defect, which dissipates with the much slower rate \cite{Dwt2}. In Ref. \cite{Dwt2} it is shown for the exponential 
model that for $r$ fixed, the outcome of the evolution 
(a rupture or a hill-and-valley structure) depends on $\epsilon_\gamma$ and the initial condition, 
i.e. the width and the depth of the pinhole. As will be seen, in the algebraic model the outcome depends also on $m$, 
which sets the rate of change of the wetting potential.
The focus is on the rate of the extension of the pinhole tip in the algebraic model, 
since the detailed computations for the exponential model are performed in Ref. \cite{Dwt2}, and morphologies are
similar in both models. 
Also, since the parameter domain of film rupture is more narrow for the algebraic model,
we investigate deep pinholes only.

The initial condition is taken as in Ref. \cite{Dwt2}, i.e. the Gaussian curve:
\begin{equation}
z(x,0) = 1-d\exp{\left[-\left(\frac{x-5}{w}\right)^2\right]},\quad 0\le x\le 10,\quad 0 < d < 1.
\label{ini_cond}
\end{equation}
Note that the length of the computational domain 
equals to ten times the unperturbed film height, and the defect is positioned at the center of the domain.
Periodic boundary conditions are used. 

Figures 5 and 6 show the log-normal plots of the pinhole depth vs time, for $d=0.9$ and $w=2,\; 0.15$, respectively.
$z_m$ is the height of the surface at the tip of the pinhole. The wide pinhole dewets for $m=1$ only. 
(See Fig. 5. Note that the exponential model predicts faster dewetting.) Wetting potentials
with $m=3$ and $m=5$ result in the quasiequilibrium at $0<z_m<1$. Quasiequilibrium means that $z_m$ (or, equivalently, the depth) changes very slow or not at all, while the rest of the shape 
changes relatively fast. In the inset, for $m=3$ one can see the onset of the formation of the hill-and-valley structure
near the endpoints of the domain; as has been noted in the Introduction, this does not affect the pinhole depth.
Interestingly, here the pinhole tip is blunt
at quasiequilibrium, while it is sharp in all examples computed for the exponential model \cite{Dwt2}. 
\begin{figure}[!t]
\includegraphics[width=3.5in]{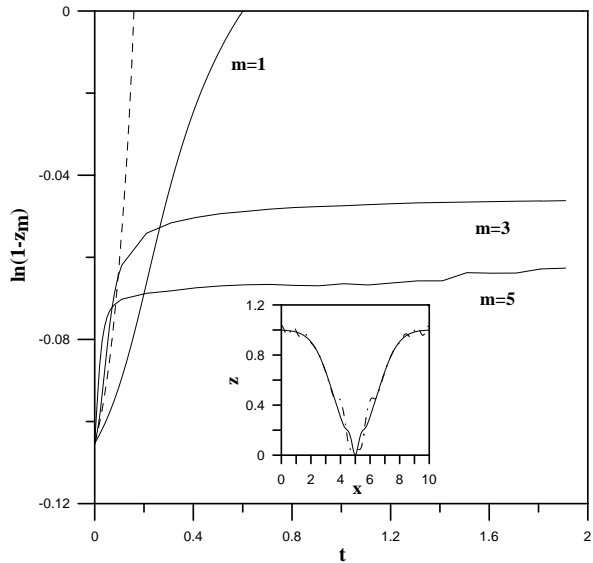}
\vspace{-2.0cm}
\caption{Kinetics (rate) data for the deep, wide pinhole ($d=0.9,\; w=2$). 
Line slope equals the rate of the tip evolution. Solid lines: algebraic model. Dash line: exponential model.
Inset:  Surface shapes at $t=1.8$, for $m=1$ (solid line) and $m=3$ (dash-dot line).} \label{fig:6}
\end{figure}
In contrast to the wide pinhole, the narrow pinhole does not dewet even for $m=1$ (Fig. 6) and 
in all three cases evolves to quasiequilibrium.
\begin{figure}[!t]
\includegraphics[width=3.5in]{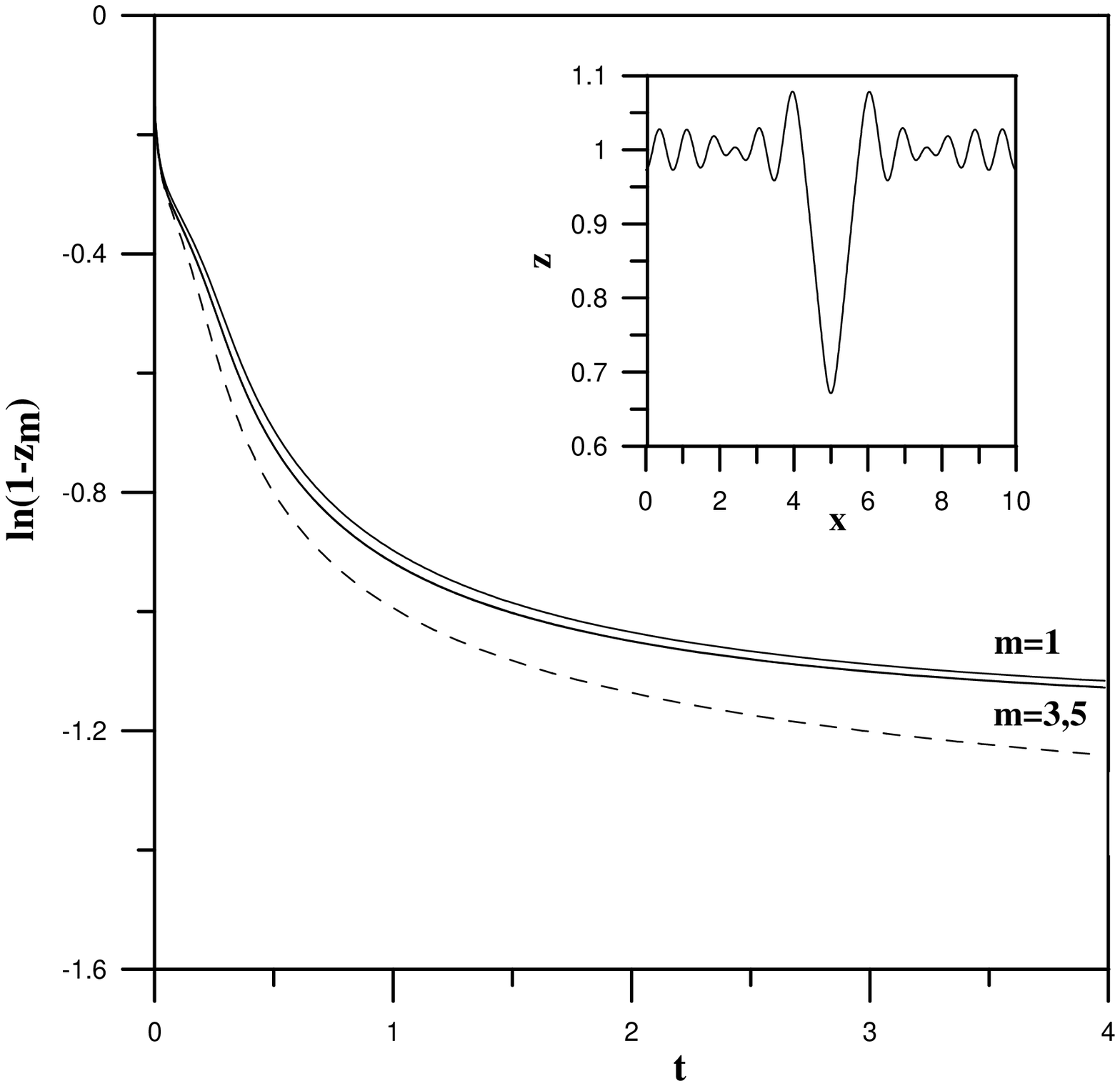}
\vspace{-2.0cm}
\caption{Kinetics data for the deep, narrow pinhole ($d=0.9,\; w=0.15$). 
Solid lines: algebraic model. Dash line: exponential model. Inset: quasiequilibrium surface 
shape at $t=4$, for $m=1$.} \label{fig:7}
\end{figure}

It must be noted here that stable equilibrium (steady state) solutions have been numerically found in the studies of a 
nonlinear stress-driven 
morphological instability of a solid film without wetting interaction, by Spencer \& Meiron \cite{SM} 
and by Xiang \& E \cite{XE}. 
The problem under study in this paper differs from the problem studied by these authors in that
the instability is driven not by stress but by wetting potential, and the surface energy is anisotropic.
These instability mechanisms have different physical origins and the process of morphological evolution
in both cases is similar but not the same.   In particular, due to the presence of strong surface energy anisotropy
the equilibrium solution, when it occurs, is replaced by quasiequilibrium. The latter can be viewed as the locally broken equilibrium. This violation of equilibrium occurs in the surface regions away from the pinhole tip. There, 
an evolving hill-and-valley structure is energetically favorable 
because the attraction to the substrate is weak. 

Finally, we note that the
slight decrease of the initial depth results in the termination of dewetting even for wide pinholes.
For instance, Figure 7 shows the case $d=0.7,\; w=2$. As can be seen, there is no dewetting for neither value of $m$.
The pinhole tip is attracted to the substrate for a while, but then reverses the direction and will finally 
stabilize at a quasiequilibrium position. Quasiequilibrium is achieved for the $m=5$ case.
For comparison, the exponential model predicts dewetting even for the more shallow pinhole with $d=0.5$
(see Fig. 2(a) in Ref. \cite{Dwt2}). 
\begin{figure}[!t]
\includegraphics[width=3.5in]{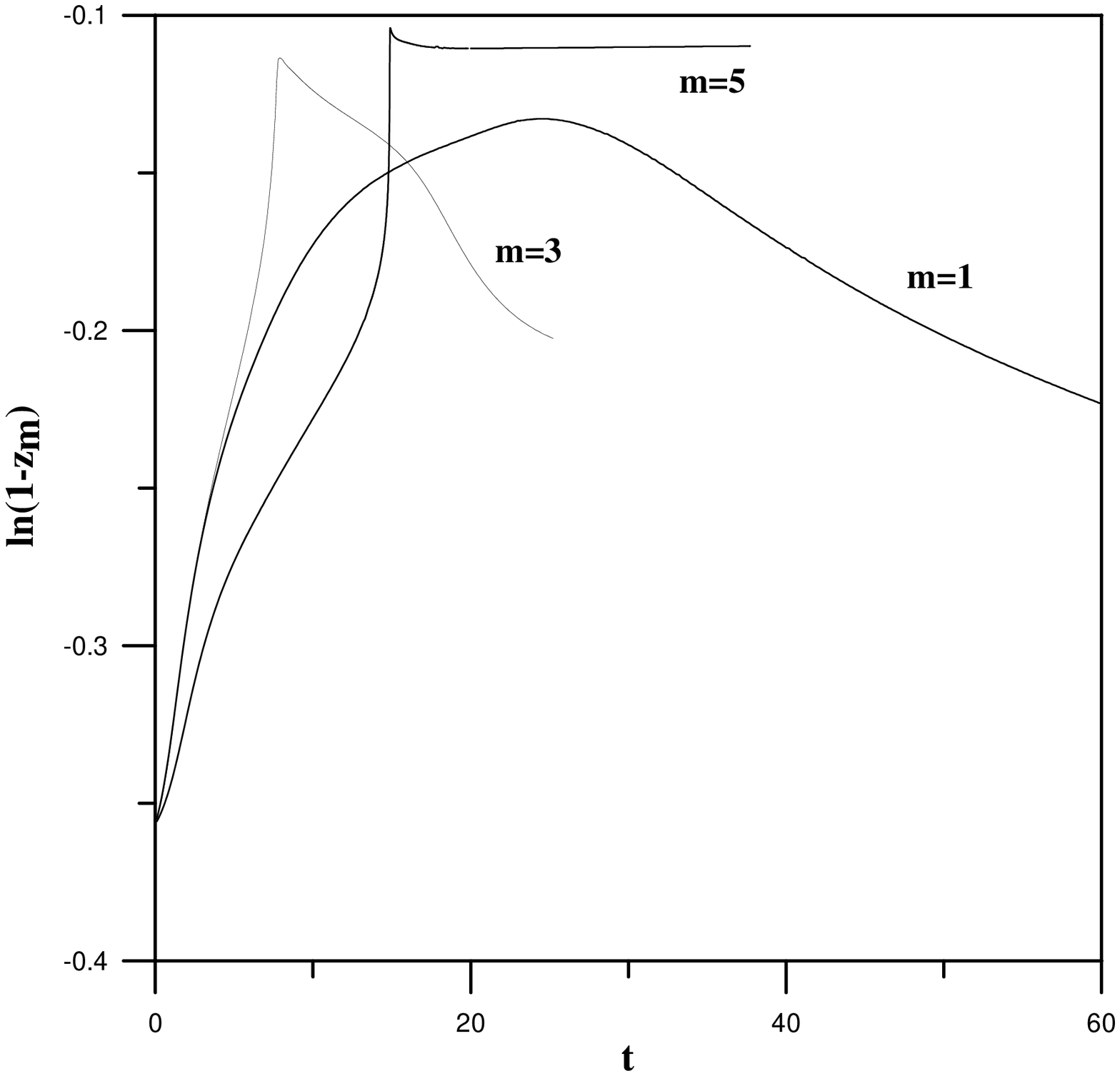}
\vspace{-2.0cm}
\caption{Kinetics data. $d=0.7,\; w=2$.} \label{fig:8}
\end{figure}
%

%\bigskip
%\begin{center}
%{\bf V. SUMMARY}
%\end{center}
To summarize, we contrasted two PDE-based models of dewetting for nonwetting ultrathin single-crystal films.
It remains to be seen how these models compare to experiment. Detailed experiments focusing
on the dynamics of a single pinhole are yet to be performed. (The published experiments \cite{YZSLLHL,SECS} 
describe very briefly the initial stages of film dewetting and proceed to detailed study of the 
post-dewetting regimes, i.e. the hole widening, secondary instabilities and material agglomeration.)

\begin{center}
{\bf ACKNOWLEDGMENT}
\end{center}

I thank Brian J. Spencer for pointing out the algebraic decay model to me.


\begin{thebibliography}{200}

\bibitem{YZSLLHL} B. Yang, P. Zhang, D.E. Savage, M.G. Lagally, G.-H. Lu, M. Huang, and F. Liu,
%``Self-organization of semiconductor nanocrystals by selective surface faceting",
\textit{Phys. Rev. B}$\;$ {\bf 72}, 235413 (2005).

\bibitem{SECS} P. Sutter, W. Ernst, Y.S. Choi, and E. Sutter,
%``Mechanisms of thermally induced dewetting of ultrathin silicon-on-insulator",
\textit{Appl. Phys. Lett.}$\;$ {\bf 88}, 141924 (2006).

\bibitem{Oron}
A.~Oron, S.~H.~Davis, and S.~G.~Bankoff, Rev. Mod. Phys. {\bf 69},
931 (1997).
%A.~Oron, S.H.~Davis, S.G.~Bankoff ``Long-scale evolution of thin liquid
%films,'' Rev. Mod. Phys. {\bf 69}, 931-980 (1997).

\bibitem{SHJ}
R.~Seemann, S.~Herminghaus, and K.~Jacobs, J. Phys.: Condensed
Matter {\bf 13}, 4925 (2001).
%``Gaining control of pattern formation of dewetting films,"

\bibitem{Dwt1} M. Khenner,
\textit{Phys. Rev. B}$\;$ {\bf 77}, 165414 (2008).

\bibitem{Dwt2} M. Khenner,
\textit{Phys. Rev. B}$\;$ {\bf 77}, 245445 (2008).
%submitted. Preprint available at http://arxiv.org/abs/0804.1816.

\bibitem{GDV} A.A.\ Golovin, S.H.\ Davis, and P.W. Voorhees, 
%``Self-organization of quantum dots in epitaxially strained solid films", 
\textit{ Phys. Rev. E}$\;$ {\bf 68}, 056203 (2003).

\bibitem{GLSD} A.A. Golovin, M.S. Levine, T.V. Savina, and S.H. Davis,
%``Faceting instability in the presence of wetting interactions: A mechanism for the formation of quantum dots ", 
\textit{ Phys. Rev. B}$\;$ {\bf 70}, 235342 (2004).

\bibitem{SVD} T.V. Savina, P.W. Voorhees, and S.H.\ Davis,  
%``The effect of surface stress and wetting layers on morphological instability in epitaxially strained films", 
\textit{J. Appl. Phys.}$\;$ {\bf 96}, 3127 (2004).

\bibitem{LGDV} M.S. Levine, A.A. Golovin, S.H. Davis, and P.W. Voorhees,
%``Self-assembly of quantum dots in a thin epitaxial film wetting an elastic substrate", 
\textit{ Phys. Rev. B}$\;$ {\bf 75}, 205312 (2007).

\bibitem{CG} C.-h. Chiu and H. Gao, in \textit{Thin Films: Stresses and Mechanical Properties V},
edited by S.P. Baker, MRS Symposia Proceedings No. 356 (Materials Research Society, Pittsburgh, 1995), p. 33.

\bibitem{Chiu} C.-h. Chiu, 
%``Stable and uniform arrays of self-assembled nanocrystalline islands", 
\textit{ Phys. Rev. B}$\;$ {\bf 69}, 165413 (2004).

\bibitem{SZ} Z. Suo and Z. Zhang,
\textit{ Phys. Rev. B}$\;$ {\bf 58}, 5116 (1998).

\bibitem{MULLINS5759} W.W.\ Mullins, 
%Theory of Thermal Grooving,
\textit{J. Appl. Phys.}$\;$ {\bf 28(3)}, 333 (1957); \textit{J. Appl. Phys.}$\;$ {\bf 30}, 77 (1959).

\bibitem{BrianWet} B.J. Spencer,
\textit{ Phys. Rev. B}$\;$ {\bf 59}, 2011 (1999).

\bibitem{KF} R.V. Kukta and L.B. Freund,
\textit{J. Mech. Phys. Solids}$\;$ {\bf 45}, 1835 (1997).

\bibitem{AG} S.\ Angenent and M.E.\ Gurtin, \textit{Arch. Rational Mech. Anal.}$\;$ {\bf 108}, 323 (1989).

\bibitem{CGP} A.\ Di Carlo, M.E.\ Gurtin, and P.\ Podio-Guidugli,
%``A Regularized Equation for Anisotropic Motion-by-Curvature",
\textit{SIAM J. Appl. Math.}$\;$ {\bf 52}, 1111 (1992).

\bibitem{BP} H.P. Bonzel and E.Preuss, 
%``Morphology of periodic surface profiles below the roughening temperature: aspects of continuum theory", 
\textit{Surface Science}$\;${\bf  336}, 209 (1995).

\bibitem{Brian_regular} B.\ J.\ Spencer,  
%``Asymptotic solutions for the equilibrium crystal shape with small corner energy regularization", 
\textit{Phys. Rev. E}$\;$ {\bf 69}, 011603 (2004). 

\bibitem{GoDaNe98} A.A.\ Golovin, S.H.\ Davis, and A.A.\ Nepomnyashchy, 
%``A convective Cahn-Hilliard model for the formation of facets and corners in crystal growth", 
\textit{ Physica D}$\;$ {\bf 122}, 202 (1998).

\bibitem{LiuMetiu} F. Liu and H. Metiu, \textit{Phys. Rev. B} {\bf 48}, 5808 (1993).

\bibitem{SG} J. Stewart and N. Goldenfeld, \textit{Phys. Rev. A}$\;$ {\bf 46}, 6505 (1992).

\bibitem{HERRING} C.\ Herring, in
\textit{Structure and Properties of Solid Surfaces},
Eds. R. Gomer and C.S. Smith (Univ. Chicago Press, Chicago, 1953)
$\;$ 5-81.

\bibitem{HERRING1} C.\ Herring, 
%``Some Theorems on the Free Energies of Crystal Surfaces",
\textit{Phys.\ Rev.}$\;$ {\bf 82}, 87 (1951).

\bibitem{Sethian1} J.A. Sethian,
\textit{Comm. Math. Phys.}$\;$ {\bf 101}, 487 (1985).

\bibitem{Sethian2} J.A. Sethian,
\textit{J. Diff. Geom.}$\;$ {\bf 31}, 131 (1990).

\bibitem{BKL} R.C. Brower, D.A. Kessler, J. Koplik, and H. Levine,
\textit{ Phys. Rev. A}$\;$ {\bf 29}, 1335 (1984).

\bibitem{HLS} T.Y. Hou, J.S. Lowengrub, and M.J. Shelley,
\textit{ J. Comput. Phys.}$\;$ {\bf 114}, 312 (1994).

\bibitem{RADAU}  E.\ Hairer and G.\ Wanner, 
%``Stiff differential equations solved by Radau method",
\textit{ J. Comput. Appl. Math.}$\;$ {\bf 111}, 93 (1999).

\bibitem{SM} B.J. Spencer and D.I. Meiron,
\textit{Acta Metall. Mater.}$\;$ {\bf 42}, 3629 (1994).

\bibitem{XE} Y. Xiang and W. E,
\textit{J. Appl. Phys.}$\;$ {\bf 91}, 9414 (2002).



\end{thebibliography}
\end{document}